\documentclass[sn-mathphys,Numbered]{sn-jnl}

\usepackage{lineno}
\usepackage{bm}

\usepackage{graphicx}%
\usepackage{multirow}%
\usepackage{amsmath,amssymb,amsfonts}%
\usepackage{amsthm}%
\usepackage{mathrsfs}%
\usepackage[title]{appendix}%
\usepackage{xcolor}%
\usepackage{textcomp}%
\usepackage{manyfoot}%
\usepackage{booktabs}%
\usepackage{algorithm}%
\usepackage{algorithmicx}%
\usepackage{algpseudocode}%
\usepackage{listings}%

\DeclareMathOperator*{\argmin}{argmin}

\title{Global Estimation of Range Resolved Thermodynamic Profiles from MicroPulse Differential Absorption Lidar}

\author*[1]{\fnm{Matthew} \sur{Hayman}}\email{mhayman@ucar.edu}

\author[1]{\fnm{Robert A.} \sur{Stillwell}}\email{stillwel@ucar.edu}

\author[1]{\fnm{Adam} \sur{Karboski}}\email{karboski@ucar.edu}

\author[2]{\fnm{Willem J.} \sur{Marais}}\email{willem.marais@ssec.wisc.edu}

\author[1]{\fnm{Scott M.} \sur{Spuler}}\email{spuler@ucar.edu}

\affil*[1]{\orgdiv{Earth Observing Lab}, \orgname{National Center for Atmospheric Research}, \orgaddress{\street{PO Box 3000}, \city{Boulder}, \postcode{80307}, \state{CO}, \country{USA}}}

\affil[2]{\orgdiv{Space Science and Engineering Center}, \orgname{University of Wisconsin at Madison}, \orgaddress{\street{1225 West Dayton St}, \city{Madison}, \postcode{53726}, \state{WI}, \country{USA}}}

\abstract{
We demonstrate thermodynamic profile estimation with data obtained using the MicroPulse DIAL such that the retrieval is entirely self contained.  The only external input is surface meteorological variables obtained from a weather station installed on the instrument.  The estimator provides products of temperature, absolute humidity and backscatter ratio such that cross dependencies between the lidar data products and raw observations are accounted for and the final products are self consistent.  The method described here is applied to a combined oxygen DIAL, potassium HSRL, water vapor DIAL system operating at two pairs of wavelengths (nominally centered at 770 and 828 nm).  We perform regularized maximum likelihood estimation through the Poisson Total Variation technique to suppress noise and improve the range of the observations.  A comparison to 119 radiosondes indicates that this new processing method produces improved temperature retrievals, reducing total errors to less than 2 K below 3 km altitude and extending the maximum altitude of temperature retrievals to 5 km with less than 3 K error.  The results of this work definitively demonstrates the potential for measuring temperature through the oxygen DIAL technique and furthermore that this can be accomplished with low-power semiconductor-based lidar sensors.}

\begin{document}

\flushbottom
\maketitle
\thispagestyle{empty}

\section{Introduction}

The MicroPulse DIAL (MPD) is designed to perform thermodynamic profiling of the atmosphere using water vapor differential absorption lidar (DIAL), high spectral resolution lidar (HSRL), and oxygen DIAL techniques.  A testbed network of five such MPD units has been developed at NSF NCAR.  While water vapor DIAL and HSRL have long legacies in lidar technology, oxygen DIAL is relatively unexplored, with the first demonstrations of the approach occurring on the MPD starting in 2020 \cite{Stillwell2020}.  The oxygen DIAL technique was definitively described in \cite{Bosenberg1998} in the late 1990s, but the approach was dismissed as a viable solution for temperature profiling due differences in oxygen absorption between light scattered by aerosol and molecules.  After scattering, aerosol returns undergo the same narrow band absorption on the return trip, but Doppler broadening from molecular scattering will result in a change in the absorbing effect.  This is commonly referred to as Rayleigh-Doppler effect in DIAL and is typically a small error term that is either negligible, if the absorption feature is sufficiently wide, or can be mitigated by operating at an inflection point on the side of the absorption feature\cite{Spath2020}.  In the case of oxygen DIAL, however, the effect of the error becomes significant in part due to the narrow width of the oxygen lines under consideration (roughly the same width as the broadening effect) and because it is necessary to operate near the peak of the absorption feature to obtain sufficient sensitivity.  Uncorrected, this can result in temperature errors on the order of 10s of degree C which \cite{Bosenberg1998} noted renders the technique relatively little value in the field of atmospheric science.

In spite of these challenges, there are some important practical benefits to performing temperature profiling using the oxygen DIAL technique.   The DIAL technique is well suited for low power instrument operation like that employed for water vapor in the MPD.  The technique is conducive to employing photon counting with narrow band filters that enable daytime operation with eye-safe (class 1M) semiconductor lasers.  In addition, it has the potential to provide observations which are internally calibrated.  Such a capability, makes the MPD one of the few instruments providing thermodynamic observations independent of radiosondes.  Sondes are often used as the gold standard thermodynamic profiling, but it is important to recognize that they have biases that have to be corrected and calibrated.  Catching errors in sondes is extremely difficult in part because they are used as the basis of calibration or priors for many retrievals and ingested into models and reanalysis \cite{Cady-Pereira2008}.  

To enable temperature measurements through oxygen DIAL, the MPD team proposed mitigating the Rayleigh-Doppler errors through an integrated HSRL, which directly measures the composition of aerosol to molecular scattering in the detection channels.  This information then supplies the knowledge needed to correct the oxygen spectroscopy and enable accurate temperature retrievals.  Thus when combined with existing HSRL and water vapor DIAL, (the oxygen DIAL's relationship to temperature is also dependent on water vapor concentration), it should be possible to measure atmospheric temperature through the oxygen DIAL technique.

While the combination of water vapor DIAL, HSRL and oxygen DIAL should carry sufficient information to estimate atmospheric water vapor, backscatter ratio and temperature, the means of estimation has remained a significant challenge.  The current method we employ relies on a serialized process, where water vapor DIAL and HSRL data are initially processed using conventional techniques to obtain absolute humidity and backscatter ratio estimates.  Those products are then inputs to a perturbative processing method \cite{Bunn2018,Repasky2019} to provide corrections in the oxygen number density and absorption spectroscopy.  While this approach has shown promise, the following represent some key drawbacks:
\begin{itemize}
    \item Noisy data products are being used to provide corrections to the highly noise sensitive oxygen DIAL retrieval, thereby compounding noise.
    \item Water vapor and HSRL retrievals have cross dependencies and dependence on atmospheric temperature, requiring assumptions to process.
    \item Some MPD channels are under utilized because there is no seamless way of integrating their overlapping information content into the standard processing approach.
\end{itemize}

In this work we address these issues by performing a full estimate of absolute humidity, temperature and backscatter ratio across all observation channels.  This is achieved by forward modeling the retrieved variables onto observations of all six MPD channels and optimizing through a maximum likelihood estimation framework while suppressing noise through total variation regularization.  The optimization approach leveraged in this work utilizes the Poisson Total Variation (PTV) technique, which has been successfully applied to HSRL \cite{Marais2016} and MPD water vapor retrievals \cite{Marais2022}.  In this case, the temperature retrieval requires significantly more retrieved products which has resulted in a need for increased PTV processing speed and methods for reducing the problem size so it can be solved with a manageable amount of time and resources.

This work describes the method we have developed for estimating range resolved atmospheric water vapor, temperature and backscatter ratio using the MPD instrument such that no external data sources regarding atmospheric state are employed other than surface data obtained from a sensor installed on the instrument.  In this way, the retrievals are completely independent of all other observations and models.

Because prior works have thoroughly addressed estimation of water vapor and aerosol backscatter, the primary focus here is on the temperature retrievals, which represent the most difficult retrieval and the technique with the shortest legacy.


\section{MicroPulse DIAL Architecture}

The MPD is a semiconductor-based photon counting lidar designed around the concept of quantitative atmospheric profiling in a package that accommodates network deployment.  The instrument is designed to be low cost, low maintenance, portable and operate with an eye-safe (class 1M) laser.  The original MPD concept started with a water vapor only design at Montana State University \cite{Nehrir2011} which was advanced and demonstrated at NSF NCAR \cite{Spuler2015, Weckwerth2016, Spuler2021}.  The instrument concept was later extended to include HSRL capability using a rubidium vapor cell as an aerosol filter \cite{Hayman2017}.  The HSRL filter was changed to potassium, to allow for reduced hardware when further merging an oxygen DIAL into the instrument for temperature measurements \cite{Stillwell2020}.

\begin{figure}[h!]
\centering\includegraphics[width=13cm]{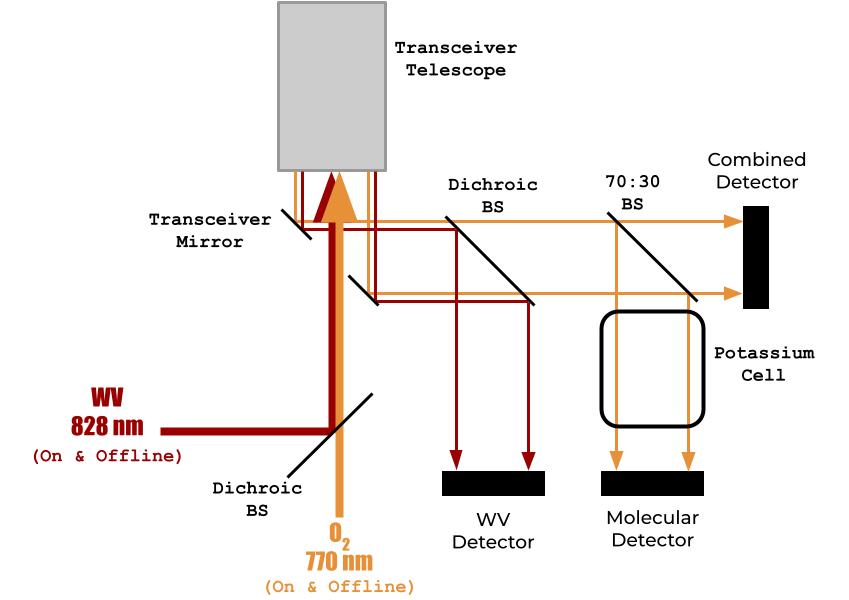}
\caption{Conceptual diagram of the MPD operating at two main wavelengths (each with alternating on- and offline laser frequency) and three detector channels resulting in a total of six unique measurements.}
\label{fig:MPD}
\end{figure}

A schematic description of the MPD, useful for understanding the formulations covered in this work, is shown in Fig. \ref{fig:MPD}.  As shown in the diagram, the MPD operates at two main wavelength bands, each alternating between closely spaced on and offline wavelengths while employing a total of three detectors.  This results in a total of six observations from the entire instrument.

The 828 nm wavelengths correspond to the water vapor (WV) DIAL system.  As is typical of DIAL, the laser output alternates between frequencies tuned on a water vapor absorption line (online) and immediately off the line (offline).  A single detector is dedicated to the 828 wavelengths.  The second set of wavelengths are near 770 nm where temperature sensitive oxygen lines and the potassium D1 transition are in close wavelength proximity to enable a combined oxygen DIAL and HSRL.  Like the water vapor DIAL, the laser frequencies are alternated between a frequency tuned to the selected oxygen absorption line and off the line.  In this case the offline frequency is tuned to the potassium D1 line to enable the HSRL functionality also integrated in the instrument.  In the receiver, the 770 nm light is split using a 70:30 beam splitter, where the larger fraction is directed through the potassium cell to block aerosol returns and provide a molecular only observation with the molecular detector (offline wavelength only).  The other 30 percent of the light is directed onto the combined detector which operates similar to the water vapor detector.  

While the oxygen offline frequency is used for the HSRL, the online frequency produces nearly redundant measurements between the molecular and combined detectors.  These two online measurements are identical within a scalar constant (due to differences in path efficiencies) and a differential overlap term caused by slight differences in detector alignment and capture of the return light.  As such, these two channels carry very similar information content with respect to the temperature profile but it is difficult to leverage both channels effectively when estimating temperature through the standard processing approach. 

The reader should note that only the concept of the MPD architecture is presented here.  A number of components have been omitted or consolidated for simplicity.  For a complete description of the MPD see \cite{Spuler2021, Stillwell2020, Spuler2015}.






\section{Forward model}

PTV employs a forward modeling approach to estimate derived products from raw photon count observations. The derived products represent the estimates that best explain the Poisson distributed, noisy observations in each channel. Regularization is then used to prevent over fitting. In this way, cross dependencies between channels can be accounted for and integrated into a single retrieval which can suppress noise.  The forward model employed here is based on the lidar equation, which includes both time, range and frequency dimensions to map the atmospheric derived variables absolute humidity, backscatter ratio and temperature along with a series of instrument terms onto the noisy observations.


Each MPD observation can be described through a general model where specific terms in a channel may differ and are denoted by superscripts.  These differences may depend on laser wavelength $^{(w)}$ (e.g. o2on for oxygen DIAL online wavelength), detector $^{(d)}$ (wv, comb or mol for water vapor, combined or molecular detectors respectively -- each detector in Figure \ref{fig:MPD}) or both $^{(c)}$ such that the term is channel specific.  For this work, the general description for the profile of a received signal as a function of range bin $k$ and relative frequency $\nu$ where we account for the Rayleigh-Doppler effect in which the narrow band laser light is broadened by molecular scattering \cite{Spath2020}.  In this work $\nu$ is referenced to the laser transmit frequency and thus may assume negative values.  As such, the frequency resolved lidar equation separates the outgoing and return narrow band absorption terms (i.e. water vapor and oxygen resonances) where the absorption experienced by molecular backscattering is different from that of aerosol backscattering.  With the assumption that the laser pulse is represented by a delta function in range and the laser spectrum is a delta function in frequency, we employ the following forward model for any MPD channel

\begin{minipage}{\linewidth}
\begin{align}\label{GeneralChannels}
    \hat{u}_k^{(c)}(\hat{\bm{n}},\hat{\bm{T}},\hat{\bm{B}},\hat{\bm{\phi}}) = &\overbrace{\,G^{(w)} C_k^{(d)} \hat{\phi}_k e^{-\omega_{0,k}^{(w)}(\hat{\bm{n}},\hat{\bm{T}})}}^{\hidewidth\text{Scene dependent terms and outgoing absorption}\hidewidth} \nonumber \\
    & \times \biggl[ \underbrace{\eta_0^{(c)} e^{-\omega_{0,k}^{(w)}(\hat{\bm{n}},\hat{\bm{T}})} \gamma^{(w)} \left( \hat{B}_k - 1\right)}_{\hidewidth\text{Aerosol scattering and return absorption}\hidewidth}+\overbrace{\sum_{\nu} \eta_{\nu}^{(c)} e^{-\omega_{\nu,k}^{(w)}(\hat{\bm{n}},\hat{\bm{T}})} \tilde{\beta}_{\nu,k}^{(w)}(\hat{T}_k)}^{\hidewidth\text{Molecular scattering and return absorption}\hidewidth} \biggr]
\end{align}
\end{minipage}
where $\hat{u}_k^{(c)}$ is the expected photon counts in range bin $k$ on channel $(c)$ per laser shot as a function of estimated variables (where estimated terms are indicated by $\,\hat{}$ ).  The differential geometric overlap function of the particular detector (relative to the combined detector) is $C_k^{(d)}$,  $G^{(w)}$ is a constant scalar accounting for absorption below the first range bin and differences in laser power, $\hat{\phi}_k$ represents the common terms across all the channels and largely accounts for broadband atmospheric extinction (e.g. from aerosols and clouds) and, for our parameterization, also captures the combined channel geometric overlap function.  The channel's frequency dependent receiver transmission is $\eta_{\nu}^{(c)}$ where the frequency index is indicated by subscript $\nu$ and the transmission at the laser wavelength is $\eta_0^{(c)}$ indicated by $\nu=0$, assuming a laser spectrum well represented by a delta function.   The frequency dependent optical depth resulting from the estimated water vapor and oxygen absorption for the specified wavelength is $\omega_{\nu,k}^{(c)}(\hat{\bm{n}},\hat{\bm{T}})$ (where when $\nu = 0$ this is only evaluated at the laser center frequency).  This optical depth depends on the column of estimated water vapor number density $\hat{\bm{n}}$ and temperature $\bm{\bm{T}}$, $\gamma^{(w)}$ is a wavelength scaling term for when this channel operates at a different wavelength than the HSRL system to account for differences in aerosol scattering resulting in a different backscatter ratio from the estimate $\hat{B}_k$ (which must be greater than 1).  The frequency normalized Rayleigh-Brillouin shape for the estimated temperature is $\tilde{\beta}_{\nu,k}(\hat{T}_k)$.  Note that the summation term collapses the laser frequency axis $\nu$ so that the forward model of the signal has only range dimensions (where the time dimension is implicitly the result of multiple observations).

The contributing terms in Eq. \eqref{GeneralChannels} may be physically understood as a series of instrument and scene dependent scaling terms with an outgoing absorption term (outside the brackets), an aerosol scattered term with return absorption (first term in the brackets) and a molecular scattered term with return absorption (second term in the brackets).

In this work retrievals are performed on 2D images of time and range to enable PTV to capture correlated structure across both dimensions.  However the formulations presented here will omit the time component for simplicity.  Terms that are range dependent are represented as vectors in bold or are individually indexed as in Eq. \eqref{GeneralChannels} in range $k$ or frequency $\nu$.  Variables without indices or bold are scalars.

Under the above instrument model, we assume that the laser wavelength can be accurately represented as a delta function in frequency which creates the need to ensure that the instrument operates within the constraints that this description is valid.

The backscatter signal from Eq \eqref{GeneralChannels} is convolved in range with the laser pulse shape $\bm{l}^{(w)}$, multiplied by the number of laser shots in the accumulation interval $m^{(c)}$ and the background photon counts are added to obtain the full estimate of the observed photon counts.  In addition we include an afterpulse baseline $\bm{a}^{(c)}$ induced by the outgoing laser pulse in the transceiver and obtained through calibration.
\begin{equation}\label{forward_model}
    \hat{\bm{y}}^{(c)} = m^{(c)} \left(\bm{l}^{(w)} * \hat{\bm{u}}^{(c)} + \bm{a}^{(c)}\right)+ b^{(c)}
\end{equation}

The optical depth $\omega$ in Eq. \eqref{GeneralChannels} depends on both oxygen and water vapor absorption and is calculated using trapezoidal integration so 
\begin{equation}\label{total_opt_depth}
\omega_{\nu,k}(\bm{n},\bm{T}) = \omega_{\nu,k-1}(\bm{n},\bm{T})+\frac{1}{2}\left[\kappa_{k-1}(n_{k-1},T_{k-1}) + \kappa_{k}(n_k,T_k)\right]\Delta r_k
\end{equation}
where $\Delta r_k$ is the range spacing between points and $\kappa_k$ is the combined extinction coefficient from both species
\begin{align}
    \kappa_{k}(n_k,T_k) & = o_{\nu,k}(n_k,T_k) + q_{\nu,k}(n_k,T_k) \\
    & = f_{o2}\left( \frac{P_k}{k_B T_k} - n_k \right) \sigma_{\nu,k}(T_k) + n_k \tau_{\nu,k}(T_k)
\end{align}
where $o_{\nu,k}(n_k,T_k)$ and $q_{\nu,k}(n_k,T_k)$ are the absorption coefficients of oxygen and water vapor respectively, $f_{o2}$ is the molar fraction of oxygen in dry air (0.211032), $P_k$ is pressure, $k_B$ is the Boltzmann constant, $\sigma_{\nu,k}(T_k)$ is the temperature dependent absorption cross section of oxygen, $\tau_{\nu,k}(T_k)$ is the absorption cross section of water vapor (with weak temperature dependence).  The pressure is determined through a hydrostatic relationship which is dependent on the surface pressure (measured by a weather station on the MPD unit) and the mean (linear) temperature profile \cite{Lente2020}.
\begin{equation}\label{PfromT2}
    P_k = P_0\left( \dfrac{\Bar{T}_k}{\Bar{T}_0}\right)^{C_g},
\end{equation}
where
\begin{equation}\label{Cgdef}
    C_g = \frac{g_0 M_{air}}{R_0 L_0},
\end{equation}
where $g_0$ is the acceleration due to gravity (9.81 m/s), $M_{air}$ is the molar mass of air (28.97 g/mol), $R_0$ is the universal gas constant (8.314459 J/mol/K) and $L_0$ is the linear fit temperature lapse rate in K/km.  To obtain the this lapse rate, the mean temperature profile $\Bar{\bm{T}}$ is parameterized as a linear fit to the PTV estimated temperature $\hat{\bm{T}}$ using the lapse rate $L_0$ and a zero range intercept $\Bar{T}_0$ such that
\begin{equation}
    \Bar{T}_k = r_k L_0 + \Bar{T}_0
\end{equation}
where $r_k$ is the range at bin $k$.  The lapse rate and zero intercept are obtained from a linear fit to the temperature profile estimate
\begin{equation}
    \left[ \begin{array}{c} \Bar{T}_0 \\ L_0 \end{array}\right] = \left[ \begin{array}{cc} \bm{1} & \bm{r} \end{array}\right]^{-P}\hat{\bm{T}}.
\end{equation}
where the superscript $^{-P}$ indicates a pseudo-inverse operation.  Both of these linear fit terms are retrieved at the same temporal resolution as the temperature profile estimate.

\subsection{Absorption and Rayleigh-Brillouin spectrum}

Frequency resolved oxygen and water vapor absorption cross sections ($\sigma_{\nu,k}$ and $\tau_{\nu,k}$ respectively) are key parameters for performing thermodynamic estimation using the DIAL technique.  In particular, $\sigma_{\nu,k}$ provides the dominant sensitivity to temperature in the oxygen DIAL technique, making it imperative that this quantity accurately and rapidly update during optimization.  It also must be differentiable with respect to temperature. In this work we employ a principal component based method for calculating the oxygen absorption, water vapor absorption and Rayleigh-Brillouin spectra \cite{Hayman2019,Binietoglou2016}. In this technique, a series of principal components are "trained" on first principle calculated spectra across a range of temperature and pressure combinations and the weights of these components are fit to polynomials that are a function of temperature and pressure. This approach is significantly faster than performing the calculations based on first principles (which are prohibitively slow for an optimization routine like PTV) and easily generalized to arbitrary spectra.  Unlike lookup table approaches, it is also analytically differentiable which is also a key feature when speed is needed in numerical optimization.


This principal component approach to spectroscopy calculations runs in contrast to the parameterized inversion method proposed in other papers on temperature estimation with oxygen DIAL \cite{Bosenberg1998,Bunn2018,Repasky2019}.  In these approaches, the following approximations are made: 1) Rayleigh-Brillouin scattering is approximated as Gaussian functions, 2) the laser wavelength is assumed to be centered on the absorption line (which shifts in pressure/altitude so is not strictly possible over a large altitude range) and 3) the absorption is unaffected by additional interfering lines (there are a number of possible interfering lines in the oxygen A-band due primarily to overtone absorption and less abundant isotopes of oxygen).  

\subsection{Estimate parameterization}\label{sub:param}

In this work, the variables we estimate using PTV represent parameterizations of the physical variables described in Eq. \eqref{GeneralChannels}.  These parameterizations are used to impose soft limits on the physical variable range, impose some amount of normalization to the retrieved variable domain or better match the retrieval basis to the expected form of the physical variable.  The notation employed here is that the total estimated state is described by a single variable consisting of several specific state variables $\bm{X} = \left \lbrace \bm{x}_{\phi}, \bm{x}_{T}, ... \right \rbrace$.  In the notation of this work, these estimated state variables have a direct relationship to the physical variables in Eq. \eqref{GeneralChannels} denoted by their subscript.  For example the common terms $\bm{\phi}$ are obtained from the estimated variable $\bm{x}_{\phi}$ through a parameterizing function (defined in the third column of Table \ref{tab:parameterizations}) $\bm{\phi}= f_{\phi}(\bm{x}_{\phi})$.  Table \ref{tab:parameterizations} also includes the name used to refer to the terms and the dimensions (time $t$ and/or range $r$) in which the variable is represented.  The formulations are such that constraints do not have to be set to avoid non-physical values in most channels.  Thus the range of variables are bounded only by the parameterization function.

\begin{table}
\caption{Estimated Variables}\label{tab:parameterizations}
\begin{tabular}{|c | c | c| c |}\hline
 \textbf{Name} & \textbf{Symbol} & \textbf{Parameterization} & \textbf{Dims}\\
 \hline
 Common Terms & $\bm{\phi}$ & $\phi_k = \exp\left(x_{\phi,k}\right)$  & $t, r$ \\ 
 \hline
 Backscatter Ratio & $\bm{B}$ & $B_k = 1 + \exp\left(x_{B,k}\right)$ & $t, r$ \\
 \hline
 WV Differential Overlap & $\bm{C}^{(wv)}$ & $C^{(wv)}_k = \exp\left(x_{C^{(wv)},k}\right)$ & $t, r$\\
 \hline
  WV Number Density & $\bm{n}$ & $n_k = x_{n,k} N_{A}/M_{H_2O}$ & $t, r$\\
  \hline
  Mol. Differential Overlap & $\bm{C}^{(mol)}$ & $C^{(mol)}_k = \exp\left(x_{C^{(mol)},k-1} + x_{C^{(mol)},k} \right)$ & $r$ \\
 \hline
 Temperature & $\bm{T}$ & $T_k = T_{k-1} + \Delta r_k x_{T,k}$ & $t, r$ \\
 \hline
 Online Gain & $G^{(o2on)}$ & $G^{(o2on)} = \exp\left(x_{G^{(o2on)}}\right)$ & $t$ \\
 \hline
\end{tabular}
\end{table}

While most of these parameterizations are straightforward we should note that the Molecular Differential Overlap is a cumulative sum operation in range where the initial value $x_{C^{(mol)},0} = 0$.  Also, the retrieval of temperature is actually obtained by estimating a localized lapse rate for each point.  This is because temperature is typically better represented as a sparse set of piecewise linear functions in range than piecewise constant functions typically obtained through PTV processing.  The variable $\Delta r_k$ is the spacing in range between each retrieved point.  The surface temperature profile $T_0$ (not to be confused with $\bar{T}_0$ which is the zero intercept of the linear fit to the temperature profile) is obtained directly from the surface station measurement on the MPD.

In the definition for WV Number Density, $M_{H_2O}$ is the molar mass of water (18.018 g/mol) and $N_{A}$ is Avogadro's number.  The forward model requires number density to calculate the lidar signals, but the retrieved variable is absolute humidity, with a domain better conditioned for optimization because typical values are not large floating point numbers.

Implicit in the parameterization step is also a resampling function such that any estimated state variable may be up-sampled to the resolution of the observations.  That is, a state variable may be estimated at lower resolution than the observations.  In this work, this is only employed for the temperature ($\bm{T}$) retrieval.  This effectively constrains the resolution of the parameter which tends to be slowly varying while the retrieval is highly noise sensitive.

\subsection{Calibrations}

In order to evaluate the forward model for each channel, the MPD requires a series of internal calibrations to obtain accurate estimates of terms in Eq. \eqref{GeneralChannels}. It's important to recognize the distinction between internally and externally calibrated instrumentation.  Under internal calibration, we aim to directly measure the terms needed in the instrument model.  As such, we employ a series of hardware, software and signal processing solutions that enable direct measurement of those specific terms.

MPD data products and calibrations are never modified to obtain agreement with an external reference.  Sonde data is highly useful in the development of the instrument but only used for validation.

External calibration methods often similarly aim to characterize instrument terms in the physical model, but are obtained by forcing agreement between the instrument retrieved product and an external reference (typically a sonde).  In practice those physical terms represent "catch-all" corrections for the instrument data product while appearing to be physically justified.  That is, if the physical term were directly measured, it would not necessarily correct the retrieval by the same amount or in the same way.  Instrument models can be oversimplified or contain assumptions that are not fully justified which inherently represent "unknown unknowns".  External calibration allows one to avoid these complexities, enabling more immediate pursuit of scientific goals at the cost of rigorous understanding of the instrument, inheriting errors from the reference and the risk of inadvertently imposing biases due to incomplete sampling of factors contributing to the calibration terms.

The principle calibrations performed on MPD consist of a receiver scan to characterize the frequency dependent transmission of each channel $\eta_{\nu}^{(c)}$ and an afterpulse calibration to obtain the baseline signal induced noise from the outgoing laser pulse $\bm{a}^{(c)}$.

The receiver scan is performed by sweeping the wavelength of a seed laser over the passband for each channel, as described in \cite{Spuler2021}.  For most channels, this captures the etalon transmission as a function of frequency while accounting for angular spread in the light passing through the receiver.  A physical model of the etalon, and potassium absorption (in the case of the oxygen offline molecular channel), is used to obtain a denoised estimate of the receiver's characteristics.  In addition to the receiver spectrum shape, this also provides the relative scaling efficiency of each detector which largely mitigates the need for any additional scaling of the receiver channels ($G^{(w)}$, the channel gain accounting for differences in collected power, is strictly laser dependent and even then, 1 for most instances).

The afterpulse baselines are obtained by covering the instrument ports to the atmosphere so that any returns can be attributed to afterpulsing effects from the outgoing pulse.  These profiles are accumulated over approximately a half hour and processed using a physical model based on \cite{Horoshko2017,Ziarkash2018} where the afterpulse is a weighted sum of decaying exponentials.  Thus the forward model used to fit to noisy afterpulse data is
\begin{align}
    \rho_k = & b + a_k \\
    \rho_k = & b + \sum_{i} c_i \exp\left( -\frac{r_k}{p_i}\right)
\end{align}
where $\rho_k$ is the expected photon counts per shot at the $kth$ range bin, $b$ is a constant background term (not used in actual lidar processing), $a_k$ is the afterpulse function employed in Eq. \eqref{forward_model}, $r_k$ is the range, $c_i$ is the exponential coefficient that is used for fitting to the noisy data and $p_i$ is the decay constant in range of the afterpulse mode.

In addition to directly calibrated quantities, we can obtain the differential overlap between the oxygen molecular and combined channels $C^{(mol)}$ through the retrieval itself.  This is possible because overlap is the only major distinction between the two channels when the laser is transmitting the online wavelength.  To simplify and accelerate the estimation process, we only estimate this quantity as needed (e.g. when other calibrations are updated) and it is generally stored as a calibration parameter.

\section{Optimization}
We employ the Poisson Total Variation (PTV) technique in order to obtain denoised estimates of atmospheric and instrument variables.  This technique is able to efficiently perform total variation regularized maximum likelihood estimation.  The technique described in the seminal work \cite{Marais2016}, was applied to water vapor retrievals in \cite{Marais2022} and its implementation is described in further detail in for high resolution estimation in \cite{Hayman2023}.  Here we provide a high-level summary of the technique.  

We aim to obtain some set of estimated state variables $\bm{X}$ (consisting of all terms summarized in Table \ref{tab:parameterizations}) which may consist of both time and range dimensions.  The state variables are related to noisy photon count observations $\bm{y}^{(c)}$ (consisting of both time and range dimensions) through a forward model $\bm{\hat{y}}^{(c)}(\bm{X})$ as described in Eq. \eqref{forward_model}.

Our goal is to obtain an estimate of $\bm{X}$ which minimizes the objective function
\begin{equation}
    \hat{\bm{X}} = \argmin \mathcal{O}(\bm{X})
\end{equation}
where the objective function consists of a negative log-likelihood term $\mathcal{L}$ and a total variation (TV) regularization term with scalar $\alpha_i$ for each estimated variable $\bm{x}_i$ in $\bm{X}$ such that
\begin{equation}\label{objective_fun}
    \mathcal{O}(\bm{X}) = \mathcal{L}(\bm{X}) +  \sum_{i}  \alpha_i ||\bm{x}_i||_{TV}.
\end{equation}

The negative log-likelihood of Poisson distributed observations is used to evaluate the fit between the forward model and the noisy observations
\begin{equation}\label{loss_function}
    \mathcal{L}(\bm{X}) = \sum_c \bm{w}^{(c)}\left[ \bm{\hat{y}}^{(c)}(\bm{X}) - \bm{y}^{(c)}\ln \bm{\hat{y}}^{(c)}(\bm{X})\right],
\end{equation}
where $\bm{w}^{(c)}$ is the channel weight which includes a mask for non-Poisson data and is described in further detail in the next section.
Note that in the validation steps described later in the regularizer search process, this same equation is used to evaluate the estimate where $\bm{y}^{(c)}$ is replaced with independent validation data.  

In the estimation process, the objective function is minimized through the Spiral-TV \cite{Harmany2012} framework with gradients calculated using fast iterative shrinkage/thresholding algorithm (FISTA) \cite{Beck2009}.

The PTV software used in previous works was based on a Cython implementation designed for CPU hardware.  We have developed an independent PTV package for this new application.  This new package is built on the PyTorch machine learning package \cite{PyTorch} and is designed to leverage GPU hardware to accelerate the gradient calculations and enable faster prototyping through the package's "autograd" feature.  In addition to moving most of the data and model onto a GPU, a CUDA kernel was developed to further accelerate the implementation of FISTA \cite{Beck2009} used for gradient calculations.  This effectively removed FISTA as a bottleneck in the processing and enabled retrievals in realistic time frames.  With this new PTV processing capability, fully processing MPD data consisting of a 24 hour period typically takes between 3 and 5 hours (depending on cluster computing queue wait times) and consumes approximately 50 GPU hours on the NSF Derecho cluster computer \cite{Derecho}.  Processing speed typically depends on the size of data being processed, so shorter time periods and lower resolutions are expected to complete in shorter amounts of time and fewer GPU hours.

\subsection{Masks and weights}

Similar to the methods described in \cite{Marais2022}, masking must be applied to photon count data that is poorly approximated as Poisson distributed.  The primary example of this is observations of clouds, where DIAL data is notorious for producing non-physical data products.  We suspect that these regions cannot be corrected using standard deadtime corrections because the photon flux is both high and rapidly varying.  As a result, the captured photon counts tend to have high error that is poorly captured in the Poisson noise model and can corrupt the retrieval of the full scene.

In previous works, we employed a binary masking scheme, however in this work we apply a continuous weighting function based on a logistic function
\begin{equation}\label{weight}
    w_k'^{(c)}(z_k) = \frac{1}{1+\exp(z_k)}
\end{equation}
where the argument of the function $z_k$ is determined based on the maximum observed count rate (determined at a capture resolution of 2 seconds) relative to a configuration set maximum count rate $\rho_{max}$ (2 MHz for this work) which is scaled by the observed standard deviation in photon counts divided by that expected for a Poisson random number (square root of the mean counts+1).  Each processed time bin is an accumulation of two second raw time bins $\Delta t$ (the typical capture resolution of the instrument) which are used to directly calculate the max, mean ($\mu_{y_k}$) and standard deviation ($\sigma_{y_k}$) of the photon counts accumulated in $y_k$ and the argument to the logistic function in Eq. \eqref{weight} is
\begin{equation}
    z_k = \frac{\max(y_{k})/\Delta t}{\rho_{max}} \frac{\sigma_{y_k}}{\sqrt{\mu_{y_k}+1}}.
\end{equation}
Through this masking scheme, data points with high backscatter and high variability (in excess of what is expected from a Poisson distribution) have reduced weight on the optimization loss function and therefore have lower priority in the optimization process.

In addition to weighting each bin, we also include a scalar weighting for the entire channel based on the available photon counts for fitting.  This was also done in \cite{Marais2022}.  This scalar channel weight is given by
\begin{equation}
w^{(c)} = \frac{1}{\sqrt{\sum_k \left(w'^{(c)}_k y^{(c)}_k\right)^2}}
\end{equation}
where the scalar weight is dependent on the non-scalar weight $\bm{w}'^{(c)}$ in Eq. \eqref{weight}.  The total weight in Eq. \eqref{loss_function} is then
\begin{equation}
    \bm{w}^{(c)} = w^{(c)} \bm{w}'^{(c)}
\end{equation}

\subsection{Regularizer search space}
PTV is built on a linear piecewise basis set, meaning the retrieved parameters are assumed to be well represented by an image composed of patches of constant value.  In the noisy observations, changes from pixel to pixel for each estimated parameter can either caused by the actual parameter fluctuations or noise. Regularization (forcing the retrieval to prefer simpler solutions) allows the retrieval to balance adding complexity to the retrieved parameter while not over fitting noise. As stated in the objective function definition in Eq. \eqref{objective_fun}, each estimated variable $\bm{x}_i$ in $\bm{X}$ (described for this specific case in Section \ref{sub:param}) has an associated regularizer $\alpha_i$ which determines the amount of total variation penalty.

The standard practice in PTV is to optimize each variable regularizer $\alpha_i$, by fully mapping the associated validation error as a function of regularizer values.  This is done by solving PTV at each regularizer (or combination of regularizer) value and projecting the forward model solution onto a statistically independent set of observations of the same scene using the loss function in Eq. \eqref{loss_function}.  Obtaining this independent validation data is possible because Poisson thinning \cite{Oh2013,Marais2017,Hayman2020} enables us to split Poisson distributed observations of one scene into two independent and identically distributed (i.i.d.) sets of observations.  As such, fitting to noise is penalized when evaluating the result against the independent validation data (using Poisson thinning to optimize tuning parameters in lidar processing is covered in more detail in \cite{Hayman2020,Marais2017,Hayman2023}).

Most previous PTV instances optimize the regularization parameters by thoroughly mapping out the regularizer space across all retrieved variables.  However, in the case described here, there are six independent variables.  Fully evaluating a six dimensional regularizer space is not practical with our current software speed and computing capabilities (the computational expense increases by the power of the search space).  In order to address this we have taken a more heuristic approach that allows us to break the search process into a series of smaller problems.  This is accomplished by performing estimation in stages to obtain accurate initial conditions (which accelerate processing) and significantly localize the regularizer search space in the final global estimation step.  At each stage, the solutions obtained from previous steps are employed either as initial conditions (if the variable is estimated at the new step) or constants in the forward model.  The breakdown of the PTV processing steps are as follows:
\begin{enumerate}
    \item Estimate the Molecular Differential Overlap $\bm{C}^{o2_{mol}}$ and Common Terms $\bm{\phi}$ by fitting to the Oxygen Molecular Online and Oxygen Combined Online channel observations.
    \item Estimate Common Terms $\bm{\phi}$ by fitting to only the Oxygen Offline Molecular channel observations.
    \item Estimate the Common Terms $\bm{\phi}$ and Backscatter Ratio $\bm{B}$ by fitting to the Oxygen Offline Molecular and Combined channel observations.
    \item Estimate the Water Vapor Differential Overlap $\bm{C}^{wv}$ by fitting to only the Water Vapor Offline channel observations.
    \item Estimate the Water Vapor Differential Overlap $\bm{C}^{wv}$ and Water Vapor Number Density $\bm{n}$ by fitting to only the Water Vapor Offline and Water Vapor Online channel observations.
    \item Estimate the scalar Online Gain $G^{(o2_{on})}$ by applying an exponentially decaying fit weight (in range) and fitting to the Oxygen Online Combined channel observations.
    \item Perform global estimation of all state variables fitting to all channel observations but freeze all regularizers except temperature to their previously determined solutions.
    \item Perform global estimation of all state variables fitting to all channel observations where all regularizers are allowed to randomly vary around the minimum obtained in the previous step.
\end{enumerate}

Note that in all steps before (7.) no more than two variables are ever estimated.  This allows us to perform a full regularizer search and obtain the optimal regularizers for those preprocess steps.

Step (1.) represents an exception case for a few reasons.  First, the common terms estimate is used purely as a proxy variable for everything other than the differential overlap term and so that specific solution is not used in any subsequent steps.  In addition this step is not run every time because it adds significant time to the processing (it turns out to be one of the slowest optimization steps).  Since the key term obtained from this step is an instrument parameter that should be stable over relatively long time periods, the step is only occasionally run and the result is stored as an instrument calibration.  Most processing runs use this calibration instead of running step (1.).

The regularizer search method applied to each of these steps also differs from the typical PTV approach.  Here, in a processing run, the regularizer value is randomly generated from a Gaussian random variable (with a mean updated based on each search step).  This approach is conducive to employing a Gaussian process search (often used for hyperparameter optimization in machine learning), however we have found that centering the search on the regularizer point with the lowest validation loss is typically more robust (but slower to converge) because it is less affected by bad optimization runs (e.g. optimization fails due to significantly excessive regularization).  For the purpose of routinely processing data, this robustness is more appealing than any potential time savings (which are lost as soon as the process fails).  

The initial conditions at the start of this optimization routine are summarized in Table \ref{tab:initial_conditions}.  Most terms are initialized as constant values.  The temperature profile is initialized as a linear lapse from the surface station temperature.  Also, the water vapor differential overlap term is typically initialized based on the relative scale of signals between water vapor and oxygen channels.

\begin{table}
\caption{Initial Conditions of Variables}\label{tab:initial_conditions}
\begin{tabular}{| c | c |}\hline
 \textbf{Name} & \textbf{Initial Condition} \\
 \hline
 Common Terms & 1.0 \\ 
 \hline
 Backscatter Ratio & 1.01 \\
 \hline
 WV Differential Overlap & system dependent constant \\
 \hline
  WV Number Density & 0 \\
  \hline
  Mol. Differential Overlap & 1.0 or prior calibration \\
 \hline
 Temperature & 9 K/km lapse from surface \\
 \hline
 Online Gain & 1.0 \\
 \hline
\end{tabular}
\end{table}




\section{Uncertainty Estimation}

Uncertainty estimation represents an important element for observational data products.  For the performance assessment presented here, we require the metric as a means of providing quality control (QC) to the data products.  Data known to be invalid based on the uncertainty obtained from the instrument observations is not included in the subsequent performance analysis.

\subsection{Bootstrapping}
In the standard MPD processing routines, we leverage a combination of Poisson thinning and bootstrapping to obtain estimates of derived parameter uncertainties \cite{Spuler2021}.  Under this method, we split each channel's photon counts into two i.i.d. scenes, process each, and calculate the square difference of the data products.  This process is repeated several times and averaged to obtain a numerical estimate of the error due to shot noise.  We have found this method consistently captures the derived product errors better than linear propagation of error.

The challenge in implementing bootstrapping with this PTV processing routine is that repeating the calculations many times becomes prohibitively expensive in processing time.  To fully encompass the impact of uncertainty, the full processing chain must be repeated along with variations in initial conditions at each bootstrapping step.  Because of the number of parameters and sensitivities, it also implies a rigorous approach would likely require more bootstrapping iterations than are typical of our water vapor and HSRL retrievals.  To provide some amount of uncertainty information, we have implemented a ``bootstrap-lite'' approach in which the final global processing step is repeated 12 times with independently thinned photon counts and random Gaussian scalars are added to the initial conditions of each state variable.  The resulting standard deviation of each data product is stored from this step to help establish errors that are essential for data QC in the data products.  While this method works relatively well with water vapor products, it is very difficult to interpret in the temperature product, sometimes reporting low errors in regions we know have high uncertainty (i.e. masked regions and regions blocked by clouds).

\subsection{Evidential Neural Net}
The bootstrap-lite approach provides some information about the potential variances of the estimated variables.  However, this result is difficult to interpret, particularly for the final temperature estimate which is obtained by range integrating the estimated lapse rate.  In order to help with the uncertainty estimation process, we turned to using an evidential machine learning technique originally proposed for performing inference on machine learning output while providing uncertainty \cite{Amini2020,Schreck2023}.


The formulation used in \cite{Amini2020} leverages a parameterization of the epistemic uncertainty (uncertainty in the mean) which allows the marginal distribution to analytically evaluate.  In their approach, the epistemic uncertainty is a scaled version of the aleatoric uncertainty (the variance and its associated uncertainty) where the scale term is predicted by the neural net.  However in this work there is no profile-by-profile capability to measure a mean error (effectively a bias) based on instrument observed parameters.  This is only possible as an ensemble analysis when comparing to a reference over many profiles.  As a result, this parameterization of epistemic uncertainty is poorly suited to our application.


In this work we modify the approach to explain all uncertainties in the temperature using the epistimic (Gaussian variance) term alone.  This avoids asserting an estimate of the mean error which cannot be predicted from the MPD data on a profile-by-profile basis and would require a composite analysis (and ultimately be an external calibration).  The result of this approach is that the error data is not truly Gaussian distributed because a persistent 1.5 K bias (seen later in Section \ref{sec:Results}) is not removed in the uncertainty estimator.  These results are nevertheless quite useful in performing data QC on the temperature estimates.

Where the original evidential technique estimates a mean and variance conditioned on a Normal Inverse-Gamma prior, our application drops the mean estimation component. We aim to estimate the variance of the temperature output by assuming that the error is Normally distributed with zero mean
\begin{equation}
    f(\Delta;\sigma^2) = \frac{1}{\sqrt{2\pi\sigma^2}} \exp\left(-\frac{\Delta^2}{2\sigma^2}\right)
\end{equation}
where $\Delta$ is the difference between the MPD temperature retrieval and a temperature reference (ECMWF ERA 5 reanalysis in this case \cite{ERA5}) and $\sigma^2$ is the distribution's variance.

In accordance with \cite{Amini2020} we then impose an inverse-Gamma prior on the variance estimate
\begin{equation}
    g(\sigma^2;\alpha,\beta) = \frac{\beta^\alpha}{\Gamma(\alpha)}\left( \frac{1}{\sigma^2} \right) e^{-\beta/\sigma^2}
\end{equation}

Applying Bayesian probability theory, we can obtain the PDF of the temperature error conditioned by the inverse-Gamma prior
\begin{align}
    p(\Delta;\alpha,\beta) & = \int_0^{\infty}f(\Delta;\sigma^2)g(\sigma^2;\alpha,\beta)d\sigma^2 \\
    & = \frac{1}{2\pi} \frac{\beta^\alpha\Gamma\left(\alpha+\frac{1}{2}\right)}{\Gamma(\alpha)}\left(\frac{2}{\Delta^2 + 2\beta}\right)^{\alpha+1/2}
\end{align}

With this PDF analytically defined, the resulting loss function for the neural net is defined as the negative log-likelihood of the PDF
\begin{equation}
    \mathcal{L}^E(\alpha,\beta) = \frac{1}{2}\ln 2\pi - \alpha \ln\beta + \left(\alpha+\frac{1}{2}\right)\ln\left(\frac{\Delta^2 + 2\beta}{2}\right) + \ln\left(\frac{\Gamma(\alpha)}{\Gamma(\alpha+1/2)}\right).
\end{equation}

In addition to the negative log-likelihood, \cite{Amini2020} employs a regularization term which we adapt to the zero mean application as
\begin{equation}
    \mathcal{L}^R(\alpha) = |\Delta|(2+\alpha)
\end{equation}
so that the total loss function in the training process is
\begin{equation}
    \mathcal{L}(\alpha,\beta) = \mathcal{L}^E(\alpha,\beta) + \lambda \mathcal{L}^R(\alpha).
\end{equation}

By leveraging this loss function, the neural network is trained with  inputs of vertical temperature profile, the standard deviation in temperature across bootstrap-lite estimates and the standard deviation of lapse rate across bootstrap-lite estimates to predict values for $\alpha$ and $\beta$ which can statistically describe the error in the temperature measurements.  Once trained, the neural network provides $\alpha$ and $\beta$ for a given profile and the variance can be calculated directly from these parameters
\begin{equation}
    \sigma^2_e = E\left[\Delta^2\right] = \frac{\beta\Gamma(\alpha-1)}{\Gamma(\alpha)}.
\end{equation}
We note that this uncertainty estimate is somewhat different from that provided in \cite{Amini2020}.  Here we obtain the variance by calculating the expected value of the second moment of $\Delta$ using $p(\Delta;\alpha,\beta)$ which accounts for spread in the prior.  In \cite{Amini2020}, they computed the expected value of $\sigma^2$ directly from the Inverse Gamma PDF $g(\sigma^2;\alpha,\beta)$.  In most instances we find the differences are small, however in some cases, accounting for the spread imposed by the Inverse Gamma function will result in larger error estimates.

The neural net used to perform the inference consists of two 512 dense layers with ReLu activation functions followed by a linear output layer.  The neural net was built and trained using PyTorch.  The inputs to the neural net consist of the vertical temperature profile, the standard deviation in temperature across bootstrap-lite estimates and the standard deviation of lapse rate across bootstrap-lite estimates.  The temperature error ($\Delta$) is calculated by comparing the MPD estimate to ECMWF ERA 5 temperature \cite{ERA5,ERA5_Hersbach}.  Training, validation and test data for the process were grouped by days to minimize temporal correlation between training, validation and test datasets which require statistical independence to fulfill their respective purposes.  The selection of days for each of these sets however, was random.

While we would argue reanalysis data is not a reasonable truth standard for state variable estimation, we claim it makes a reasonable standard when performing uncertainty estimates, where the accuracy requirements are more relaxed.  For the purposes of this work, the uncertainty is strictly used as a quality control tool to determine where data should be masked.  As a practical consideration, the use of reanalysis enables us to obtain a much larger training dataset than would be possible with the relatively few sondes.  This large dataset is essential for training neural nets.

\begin{figure}[h!]
\centering\includegraphics[width=13cm]{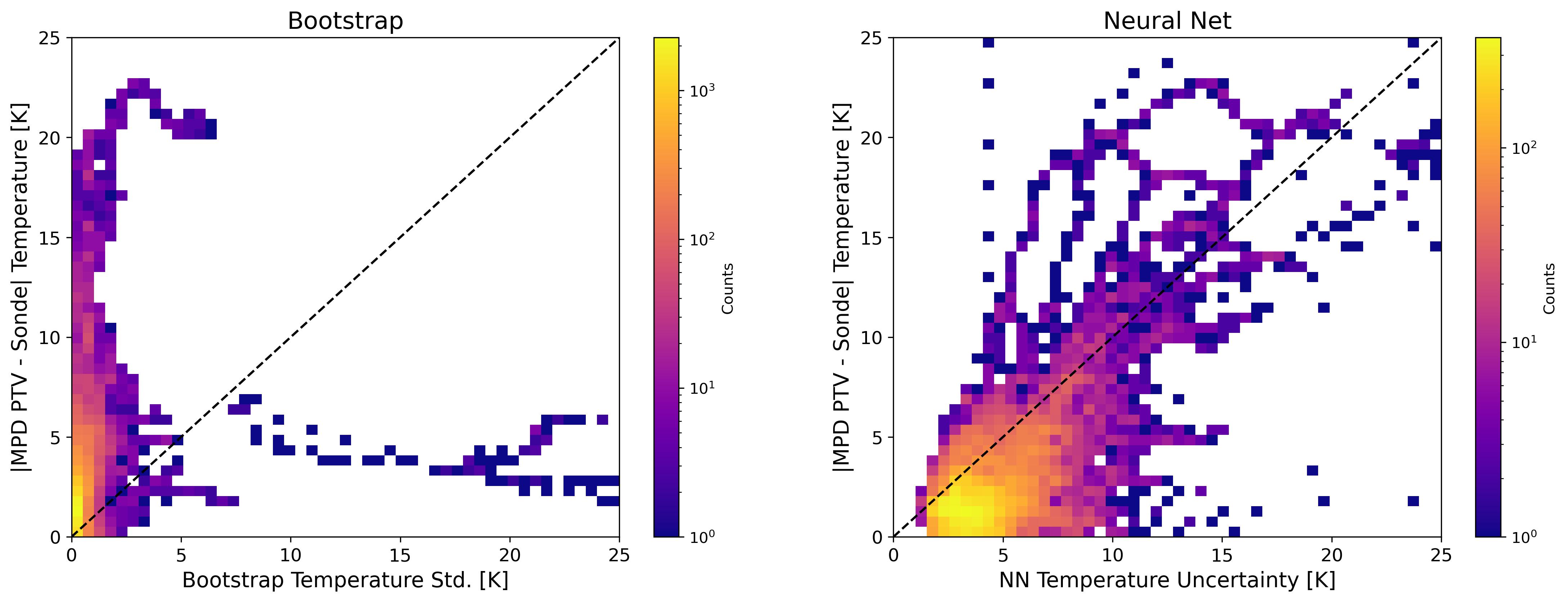}
\caption{Comparison of bootstrap standard deviation (left) and evidential neural net uncertainty (right) to temperature difference between PTV processed MPD data and sondes.  The dashed line is the 1:1 line.  Note that the bootstrap estimates are used as inputs to the neural net. The color scales are logarithmic.}
\label{fig:temperature_uncert}
\end{figure}

An analysis of the uncertainty estimation was performed by comparing the predicted uncertainties to the difference between MPD temperature estimates and 119 sonde measurements from the M2HATS field project \cite{sonde_m2hats}. Note that none of these sondes were used in the training, validation, or test process of the neural net.  The histogrammed results of that analysis for the bootstrap-lite process (described in the previous section) and the evidential neural net (which uses the bootstrap-lite process as an input) are shown in Fig. \ref{fig:temperature_uncert}.  Notably the bootstrap-lite output alone has poor correlation to the MPD/sonde differences.  The evidential neural net results, while far from perfect, represent a considerable improvement in interpreting bootstrap results with a clear correlation to MPD/sonde differences.  The performance of the evidential neural net is sufficient to perform data QC analysis critical for evaluating the performance of the retrieval method in the next section.

We would note that while this method works adequately well for temperature uncertainty estimates, using reanalysis for water vapor error estimates is not likely to be as productive because reanalysis water vapor products tend to be much less accurate than the temperature products.  For that purpose we are forced to rely directly on the bootstrap-lite results which, fortunately, are a more direct predictor of uncertainty in the water vapor product.

\section{Results}\label{sec:Results}

NSF NCAR has recently developed a testbed network of five MPD units that all measure temperature, absolute humidity and calibrated aerosol backscatter.  From July 23 to Sept 24, 2023, MPD 3 was deployed to Tonopah, NV, USA as part of the M2HATS \cite{mpd_m2hats} field campaign.  During that time, radiosondes \cite{sonde_m2hats} were launched at the MPD site twice a day typically near 1700 and 2200 UTC.  Data from the unit was processed for the full time period to obtain temperature, absolute humidity and backscatter ratio.  We applied both the standard and PTV processing approaches to the collected data and used the radiosonde products to evaluate the two approaches.

The PTV data processing employed the global thermodynamic estimation approach described in this work.  It was conducted at a base resolution of five minute time bin intervals with 37.5 m range bins.  All products except temperature were obtained at these resolutions.  Temperature, due to the high noise sensitivity, was estimated at a resolution of 40 minutes and 150 m.

The standard thermodynamic processing utilized standard water vapor DIAL \cite{Spuler2021,Spuler2015,Fix2010,Bosenberg1998} and HSRL \cite{Hayman2017, Piironen1994} processing to obtain absolute humidity and backscatter ratio estimates.  The results of those processing outputs were then supplied to the perturbative processing method described in \cite{Bunn2018,Repasky2019} which treats these products as constants in the temperature estimation process.  Temperature retrievals were processed at a base resolution of 1 minute and 37.5 m then smoothed in each perturbative step to 20 minutes and 300 m.  Finally a 40 minute median filter was applied to the data.  Uncertainty was assessed using the bootstrapping method, but importantly only applied to the temperature processing step, so uncertainties in the retrieved backscatter ratio and absolute humidity that compounded in the temperature estimate were not accounted for.

\begin{figure}[h!]
\centering\includegraphics[width=13cm]{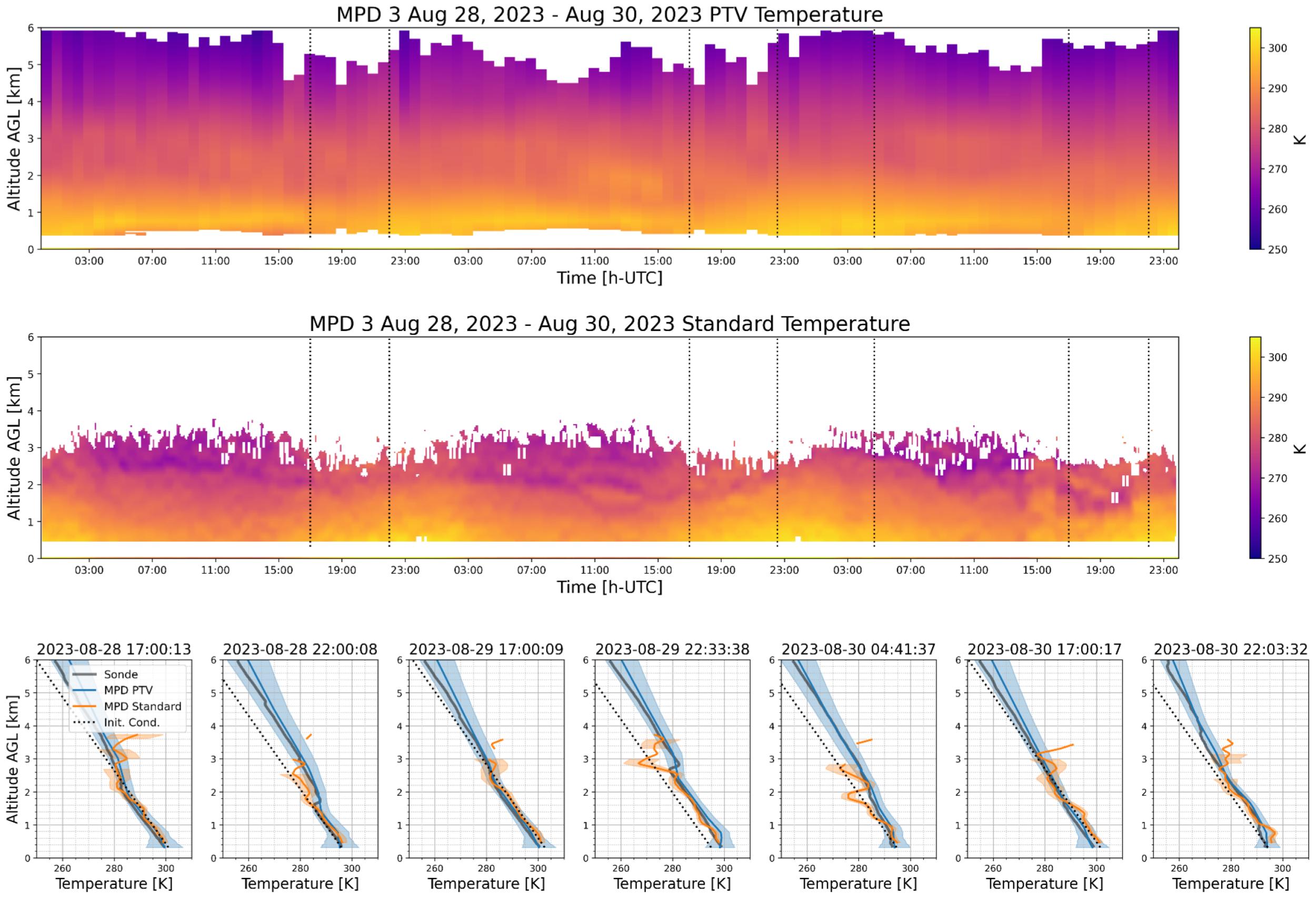}
\caption{Curtain plots of temperature estimates from August 28-30, 2023 using PTV processing (top) and standard processing (middle) with sonde times indicated by dashed lines.  The bottom row shows the temperature comparisons of the two methods (blue: PTV, orange: standard method) with the sondes (heavy gray) as well as the initial conditions (dotted black) used for the PTV retrieval.}
\label{fig:temperature_results}
\end{figure}

An example of temperature estimates from both methods is shown for a three day period spanning August 28 through August 30 in Fig. \ref{fig:temperature_results}.  The data is masked based on uncertainty analysis employed by the two techniques (bootstrapping for the standard method and evidential NN for PTV) and times of radiosonde launches are shown with vertical dashed lines.  This time period was selected because it contains a radiosonde launched near 0440 UTC, the only sonde launched at night.  Comparisons of the MPD temperature retrievals to those sondes are shown below the temperature curtain plots where the shaded regions indicate uncertainty for the respective methods determined in accordance with the aforementioned techniques.  In addition, those sonde comparisons also show the initial condition used in the PTV temperature retrieval as black dotted line (obtained by assuming a constant lapse rate of 9 K/km from the surface station).  This allows us to see that the retrieval is deviating from the initial condition and not just a product of a lucky initial guess.

Note that in this example, the PTV retrieval of temperature appears to be responsive to inversion layers.  This is important because temperature retrievals often lack features that significantly distinguish them from an assumed linear lapse rate (such as the initial condition in the optimization routine).  While the retrieval lacks sufficient resolution to track sharp changes such as that seen in 2023-08-29 22:33:38 near 3 km, it does follow the overall trend of the change.  The PTV retrieval is less responsive to noise than the standard method, generally making these changes easier to recognize and enabling the retrieval to extend high enough to capture them. 

\begin{figure}[h!]
\centering\includegraphics[width=12cm]{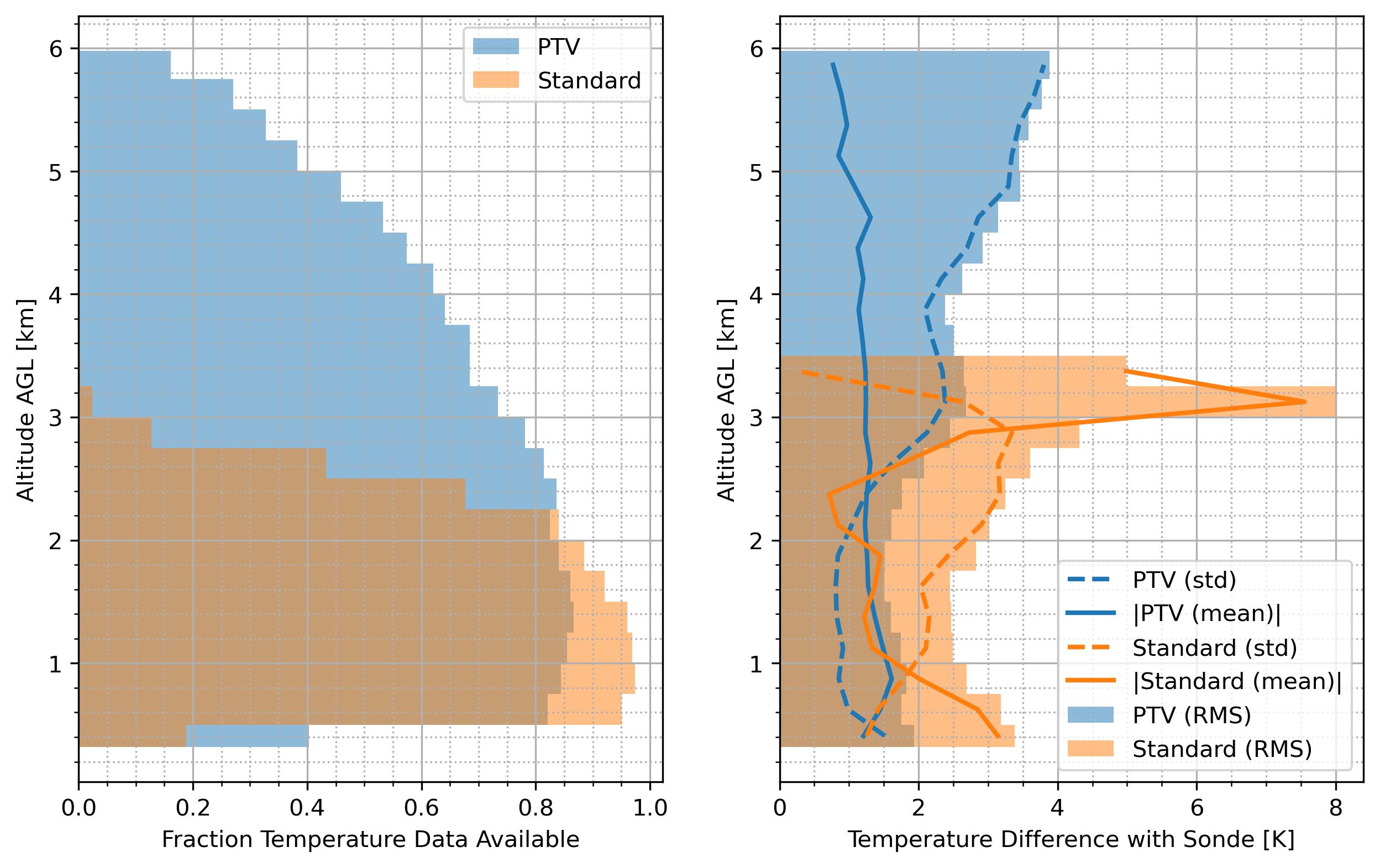}
\caption{Evaluation of MPD PTV (blue) and Standard (orange) Temperature retrievals based on data availability and agreement with sondes from M2HATS campaign.}
\label{fig:temperature_diff}
\end{figure}

We conducted an analysis of both temperature retrieval methods based on their agreement with 119 radiosondes from the M2HATS project \cite{sonde_m2hats}.  The results of that analysis are summarized in Fig. \ref{fig:temperature_diff}.  For each processing approach we only considered data that was not flagged due to high uncertainty.  In order to penalize overzealous masking, we report the data availability as a function of altitude for all sonde comparisons.  We calculate RMS difference with sonde temperature to provide a full analysis of the retrieval errors, mean difference to estimate bias in the retrievals and standard deviation to estimate spread in the retrieval errors.  We should note that all sondes, except one, were launched during the day, which represents the highest noise conditions for the lidar.

We can see that PTV tends to have slightly lower data availability in regions where the standard method is consistently valid.  This is consistent with results we reported in \cite{Marais2022} and is the result of masking due to high photon rates (particularly near clouds).  For PTV to perform accurate retrievals it requires that the Poisson noise model accurately describes the photon detection process.  However in regions of high photon flux, this noise model is no longer valid.  In order to avoid such cases corrupting the retrievals (as it impacts the entire scene, not just local to the problem area), we are forced to aggressively mask clouds and other high flux regions.  This adversely impacts the data availability.  This issue highlights the need for noise models that are accurate across larger ranges of photon flux and observational conditions.  If a noise model existed that were able to encapsulate nonlinear detection processes with heterogeneous targets, no such masking would be required to process these scenes and PTV would have higher data availability.

The temperature error in the PTV processing method is generally lower than the standard method across all overlapping altitudes (300 m to 3.5 km).  We note that the mean error tends to agree relatively well between the two methods (where there are significant data points in the standard method).  This bias of approximately 1.5 K appears to be a result of a bias in the instrument itself.  It could be attributed to errors in the measured instrument wavelength, laser spectral purity, or other aspects of the hardware that are either inaccurately captured or modeled in the processing routines.  It represents an area of ongoing development and research for the MPD team.  We note that this bias is the limiting factor in the performance of PTV retrievals below 2.4 km (where the standard deviation drops below 1 K).  This indicates that if the bias in the temperature estimate can be eliminated, the MPD PTV temperature error would significantly improve.  By contrast, the standard method error still tends to be dominated by the standard deviation for most of the valid altitude range, so reduction in biases would have a less significant impact on the data product performance.

Overall, this analysis suggests that the global estimation of thermodynamic variables using PTV produces lower data product error compared to the standard method at the cost of slightly lower data availability below 2.2 km.  However PTV also delivers much higher data product availability above 2.2 km.  Importantly PTV seems better suited to capture temperature inversions in the atmosphere, where the standard method struggles due to (typically) lower maximum range and noise which is often difficult to distinguish from atmospheric structure.

\section{Conclusion}

We have developed and demonstrated an advanced statistical signal processing approach to retrieving thermodynamic profiles from MicroPulse DIAL observations without the use of external data for processing or calibration.  The processing technique is an adaptation of PTV which has previously been demonstrated for HSRL and water vapor DIAL processing.  In this work we have introduced a temperature processing component leveraging the oxygen DIAL technique.  By using GPU-accelerated processing on a cluster computer, we are able to process a full day of data in approximately four hours at a cost of 50 GPU hours.  

A comparison to sondes shows the PTV temperature estimates are an improvement over the standard method.  The approach not only extends the range of the retrieval beyond the standard method but also reduces the errors –– to a limit imposed from a bias in the hardware itself.

While observing temperature with oxygen DIAL is a novel concept in lidar remote sensing, this work demonstrates its potential viability.  Through advances in signal processing and continued improvement in hardware performance we expect continued improvement in the temperature data products.  This is an important step toward network-capable temperature monitoring using lidar, as oxygen DIAL is well suited to low-power (eye-safe) semiconductor laser-based instrument architectures that offer high reliability and low maintenance.


\section*{Acknowledgements}
This material is based upon work supported by NSF AGS-1930907 and the NSF National Center for Atmospheric Research, which is a major facility sponsored by the National Science Foundation under Cooperative Agreement No. 1852977.  We would like to acknowledge high-performance computing support from Derecho \cite{Derecho} provided by NCAR's Computational and Information Systems Laboratory (CISL), sponsored by the National Science Foundation.

\section*{Disclosures}
The authors declare no conflicts of interest.

\bibliography{main}






\end{document}